\documentclass[reprint, 11pt,superscriptaddress,preprintnumbers,nofootinbib,aps]{revtex4}

\usepackage{graphicx}
\usepackage{amsmath,amssymb,amsfonts,amsthm,stmaryrd,mathtools,mathbbol,bm,physics,tensor}
\usepackage{color}
\usepackage{tikz}
\allowdisplaybreaks[1]

\usepackage[bookmarks,linktocpage, colorlinks=true, plainpages = false, citecolor = treegreen,  linkcolor=darkblue, urlcolor = darkblue, filecolor = blue]{hyperref} 

\def\be#1\ee{\begin{align}#1\end{align}}

\def\ba{\begin{eqnarray}}
	\def\ea{\end{eqnarray}}

\definecolor{darkblue}{rgb}{0., 0.4, 0.8}
\definecolor{cadmiumred}{rgb}{1., 0., 0.22}
\definecolor{treegreen}{rgb}{0., 0.7, 0.3}
\definecolor{orchid}{rgb}{0.7., 0., 0.5}

\begin{document}

\title{Teleocosmology and quantum post-selection}

\author{Paul C. W. Davies}
\email{Paul.Davies@asu.edu}
\affiliation{
The Beyond Center for Fundamental Concepts in Science,
Arizona State University, Tempe, AZ 85287, USA}
\author{Jo\~ao Magueijo}
\email{j.magueijo@imperial.ac.uk}
\affiliation{Abdus Salam Centre for Theoretical Physics, Imperial College London, London SW7 2AZ, UK}

\begin{abstract}
	\bigskip
	{\sc Abstract:}
Although it is widely accepted that the expansion of the universe is accelerating, the mechanism underpinning it, usually described as dark energy, remains contentious. Most explanations appeal to either a non-zero vacuum energy (i.e. a cosmological constant) or various novel fields.  In this paper we offer an alternative explanation for the cosmological acceleration stemming purely from quantum mechanics itself, without any additional physical constants or fields. The possibility arises because of the basic property of quantum indeterminism, which permits one to postulate both initial and final conditions on a quantum state--in the case considered here, that being the “wave function of the universe”. The study of quantum systems subject to both pre- and post-selection in laboratory physics is long-established, both theoretically and experimentally, and we here extend the subject to cosmology, applying it to a minisuperspace model of quantum cosmology. First we illustrate the basic idea of quantum post-selection mimicking a force, by considering the motion of a free non-relativistic particle: conditioning an initially semiclassical wave packet on a final quantum state can make the intermediate peak accelerate, even though the underlying dynamics remains free. A semiclassical observer of this phenomenon at an intermediate time would regard the acceleration as preposterous if attributed to a classical force. Such a phenomenon might then be regarded as evidence for quantum post-selection. 
We implement the analogous construction in minisuperspace quantum cosmology, using connection variables and unimodular time.  A forward semiclassical packet (describing pure radiation for simplicity) is post-selected by a normalizable Chern--Simons soliton. The resulting two--boundary amplitude has a peak which leaves the radiation trajectory and enters an accelerating regime, while the forward Hamiltonian has $\Lambda=0$. 
A classical model can mimic this trajectory only through a contrived effective component, resembling  $w\simeq -1$  near the transition and evolving towards strongly ``phantom'' behaviour ($w<-1$) when extrapolated. The acceleration is therefore more naturally interpreted as a quantum boundary-condition effect rather than as a local classical source. We close by discussing the rationale for post-selecting with a Chern--Simons soliton and possible tell-tale signatures of the model.
\end{abstract}

\maketitle
\tableofcontents

\section{Introduction: post-selection and pre-cognition}

It has been noted that, for closed systems such as the universe, quantum
mechanics may be more naturally formulated in terms of both initial and final
boundary conditions~\cite{Aharonov1964,Aharonov2008}. In such a framework, probabilities are conditioned on both
endpoints, and cosmological evolution is more appropriately viewed as a
boundary-value problem rather than a purely initial-value problem. The purpose
of this paper is to explore a concrete implementation of this idea in
minisuperspace, using for final states normalizable Chern--Simons states, and
to clarify how such boundary data feed into effective cosmological dynamics.
Relevant literature includes
\cite{Alexandre:2022npo,Gielen:2022dhg}.

The ultimate purpose is to obtain acceleration without inserting a cosmological
constant into the semiclassical Hamiltonian, and hence without inheriting its
associated fine-tuning problem (for related literature see~\cite{Anastopoulos2024}). Previous approaches to the cosmological
constant problem include mechanisms in which global classical constraints
effectively know about the spacetime history as a whole, as in vacuum-energy
sequestering
\cite{Kaloper:2013zca,Kaloper:2014dqa,Padilla:2015aaa}.
In this sense they involve a form of classical pre-cognition. Here we replace
classical pre-cognition by quantum post-selection, re-examining the same issues. 

The mechanism is simple in outline. A forward-evolved semiclassical state
defines a classical-looking trajectory without acceleration. A backward-evolved final state supplies
an additional amplitude bias. The intermediate, conditioned state is obtained
from the product of the two amplitudes and it may remain semiclassical. However, if the relative widths of the two states evolve differently, its peak motion need not coincide with the
classical trajectory of the forward state. The selected peak can 
accelerate. Importantly, this acceleration is not produced by adding a local
force, a new fluid, modified gravity or a cosmological constant to the Hamiltonian. It is
produced by the final boundary condition alone. A classical theory engineered to
reproduce the same peak motion can be concocted, but it may be abstruse.

We first demonstrate this point in the elementary case of a non-relativistic free particle. An
initial semiclassical Gaussian packet evolves according to the free Hamiltonian.
If it is conditioned on a final quantum state (for example a delta function at a position not intersecting the classical trajectory), the peak of the intermediate
conditional distribution is displaced from the free classical trajectory. In a
suitable regime the displacement is large enough to produce acceleration while
the packet remains semiclassical. The corresponding classical force required
to reproduce the same peak motion would follow an esoteric time-dependent and velocity-dependent law, hard to motivate physically. This model has an interest in its own right, and we believe an experimental verification would be clarifying. For the purposes of this paper it provides a perfect analogy for what we shall seek in cosmology. A simple quantum boundary condition can
look, to a semiclassical observer, like an awkward classical modification of
the equations of motion.

To implement the idea in cosmology we use the quantum-unimodular construction
of \cite{Alexandre:2022npo,Gielen:2022dhg,Magueijo:2021pvq}. In this
framework the cosmological constant is a priori treated as a variable, while
its conjugate provides a physical time \cite{Magueijo:2021rpi}. Classically,
the constancy of \(\Lambda\) is then an equation of motion, derived from the independence of the Hamiltonian from this physical time. Quantum
mechanically, however, one may superpose different values of \(\Lambda\). This allows
the construction of normalizable wave packets centered on a given $\Lambda_0$ (namely zero), but also of normalizable states
without a sharply defined cosmological constant. The same construction may be
expressed either in metric variables, where it is naturally related to
Hartle--Hawking states, or in connection variables, where the corresponding
states are Chern--Simons states (also called Kodama states~\cite{Kodama:1988,Smolin:2002,Magueijo:2020ugp,Alexander:2022ocp}). We use the latter representation,
since the relevant states take a particularly simple form. A further benefit of the unimodular formulation is that the Wheeler--DeWitt equation is recast as a Schrödinger-like equation, with unimodular time playing the role of the evolution parameter.

The forward state will be a semiclassical radiation-dominated universe, written
in connection space and evolved using the unimodular time $T$ conjugate to
$\phi=3/\Lambda$. (For simplicity we will ignore the matter epoch in this toy model.) 
We recall that the connection variable $b$, which on-shell satisfies $b=\dot a$ (i.e., the inverse comoving Hubble length), decreases for a decelerating
universe (for example a radiation-dominated one) but increases for any accelerating universe. Hence,  a turn-around in
$b(T)$ is the relevant signal of the onset of acceleration. We take the final state to be
a normalizable Chern--Simons soliton: a quantum state localized in the connection $b$, but with no definite value of $\Lambda$. Such a state is natural on dimensional grounds, but it is not a state with a
definite value of $\Lambda$; rather, it has no well-defined semiclassical
cosmological constant.

The central
claim is that the product of the forward radiation packet and this
backward-evolved final state can have a peak which turns around and enters an
accelerating regime, even while the forward Hamiltonian remains arbitrarily
close to that of a $\Lambda=0$ radiation universe. 
Because the soliton has no internal beats, it does not contaminate the phase
of the forward peak; the latter can therefore remain associated with
$\Lambda=0$.
A classical Friedmann description of the selected peak can of course be
reconstructed. Near the transition it resembles an effective dark component
with \(w\simeq -1\), as current observations would suggest,  but when extrapolated, the effective component evolves towards strongly phantom behaviour before eventually ceasing to be a natural local source. This mirrors the free-particle example: the quantum
description is simple, whereas the classical dynamics required to mimic it is
contrived.

The plan of the paper is as follows. In Section II we present the free-particle
model and show explicitly how post-selection produces an accelerated
semiclassical peak while the underlying dynamics remains free. In Section III
we set up the corresponding minisuperspace quantum cosmology, reviewing the
classical radiation and Chern--Simons trajectories and the associated wave
packets in unimodular time. In Section IV we construct the post-selected
cosmological peak and show that it can depart from the radiation trajectory and
turn around without introducing a cosmological constant into the forward
Hamiltonian. We also discuss why the equivalent classical Friedmann
reconstruction is unnatural. 
Section~\ref{Sec:FinalState} discusses the rationale for the Chern--Simons soliton final state,
the small overlap between initial and final states, and the role of possible
alternative final states. Section~\ref{Sec: discussion} turns to the physical interpretation,
including the distinction between laboratory post-selection and cosmological
final boundary conditions.

\section{The free particle case: acceleration from post-selection}\label{Freeparticel}

Consider a non-relativistic free particle of mass $m$. Take an initial semiclassical Gaussian wavepacket at $t=0$, with momentum $p_0\neq 0$:
\begin{equation}
\psi_i(x,0)\propto \exp\left[-\frac{(x-x_0)^2}{4\sigma^2} + i p_0 x\right].
\end{equation}
At time $t$ this becomes a Gaussian wave packet with: 
\begin{equation}
|\psi_i(x,t)|
\propto
\exp\left[-\frac{(x-x_{\rm cl}(t))^2}{4\sigma^2}\right].
\end{equation}
peaked on the classical trajectory:
\begin{equation}
x_{\rm cl}(t)=x_0+\frac{p_0}{m}t.
\end{equation}
For simplicity we have ignored the spreading of the forward packet (but see Appendix~\ref{Ap:details1} for an evaluation of this approximation).

Let the final state at $t_f$ be entirely non-classical, say a delta function $\delta(x-X_f)$, approximated as a narrow Gaussian,
\begin{equation}
\psi_f(x)\propto \exp\left[-\frac{(x-X_f)^2}{4\Sigma^2}\right].
\end{equation}
Backward evolution to time $t$ results in another Gaussian with width
\begin{equation}
W^2(\tau)=\Sigma^2+\frac{\hbar^2\tau^2}{4m^2\Sigma^2}, \qquad \tau=t_f-t,
\end{equation}
so that
\begin{equation}
|\tilde{\psi}_f(x,t)|
\propto
\exp\left[-\frac{(x-X_f)^2}{4W^2(\tau)}\right].
\end{equation}

The two-boundary amplitude at time $t$ is therefore the product
\begin{equation}
|\Psi_{\rm cond}(x,t)|
\propto
\exp\left[-\frac{(x-x_{\rm cl}(t))^2}{4\sigma^2}\right]
\exp\left[-\frac{(x-X_f)^2}{4W^2(\tau)}\right],
\end{equation}
which is again Gaussian, with peak at
\begin{equation}
X(t)=\frac{W^2(\tau)x_{\rm cl}(t)+\sigma^2 X_f}{W^2(\tau)+\sigma^2}.
\end{equation}
Writing
\begin{equation}
X(t)=x_{\rm cl}(t)+\delta X(t),
\end{equation}
one finds
\begin{equation}
\delta X(t)=\frac{\sigma^2}{\sigma^2+W^2(\tau)}\left(X_f-x_{\rm cl}(t)\right).
\end{equation}
Thus the peak lies between the classical trajectory and the location preferred by the final state, weighted by their respective widths.

At early times (specifically when $\tau\gg 2m\Sigma\sigma/\hbar$), we have $W^2(\tau)\gg\sigma^2$ and so $X(t)\approx x_{\rm cl}(t)$. At late times ($\tau\to 0$), the peak is pulled towards $X_f$. Differentiating shows that the peak obeys a complicated effective equation of motion of the form
\begin{equation}\label{accelgeneral}
\ddot X(t)
=
\frac{2\sigma^2\beta_0}{D(\tau)^3}
\left[
\beta_0 v\tau^3
+
3\beta_0\Delta\tau^2
-
3v(\sigma^2+\Sigma^2)\tau
-
\Delta(\sigma^2+\Sigma^2)
\right].
\end{equation}
A derivation and explanation of all symbols can be found in Appendix~\ref{Ap:details1}. The point is that this is equivalent to a complicated time-dependent, velocity-dependent effective force.

Thus, even though the underlying dynamics is that of a free particle, the peak exhibits an effective acceleration induced purely by the final boundary condition. Moreover, {\it this acceleration has a form that would be difficult to motivate from first principles as a classical force}. Hence, even if we only had access to the semiclassical stage of the particle's evolution, we would infer that it is conditioned by a quantum final state; any purely classical explanation would be baroque. This provides a perfect metaphor for the quantum cosmology we are about to propose.


Several consistency checks on this example are collected in Appendix B. There we show that the
conditioned packet can remain semiclassical even when the displacement of its peak is significant,
and that the phase of the two-boundary state encodes the corresponding peak momentum. We
also spell out in Appendix~\ref{energyaccounting} the energy accounting. 
Post-selection acts here as a quantum Maxwell
demon for the first law. It does not modify the free Hamiltonian of the full ensemble, but it can
select a subensemble in which the particles have genuinely gained energy relative to the
unconditioned free evolution. A classical observer restricted to that subensemble would infer an effective violation of energy conservation, though, as stressed above, 
this is merely the
equivalent classical reinterpretation, not the underlying quantum description.
The price is paid by discarding the overwhelming majority of runs. Thus the energy gain is a real
``quantum miracle'' as a property of the post-selected ensemble.

\section{Moving on to cosmology}\label{QC}
%

Albeit benefiting from the simplifications afforded by symmetry reduction, quantum cosmology inherits two inter-related problems from full quantum gravity:
the problem of time and that of defining a suitable inner product and Hilbert space\footnote{They intermingle, for example, in issues such as defining unitary evolution.}.
We do not want to wed our considerations to any particular scheme for addressing these issues, but for concreteness we will use the construction in~\cite{Magueijo:2021pvq,Alexandre:2022npo,Gielen:2022dhg}. In this framework, physical time $T$ appears as the conjugate of the constants of nature~\cite{Magueijo:2021rpi} demoted to constants of motion, for example the cosmological constant, as is the case with unimodular gravity~\cite{unimod1,Kuchar:1991xd,unimod,UnimodLee1,alan,daughton,sorkin1,sorkin2}. This converts the Wheeler-DeWitt equation into a Schr\"odinger-like equation in $T$, suggests a natural inner product and a notion of unitary evolution~\cite{Magueijo:2021pvq,Alexandre:2022npo,Gielen:2022dhg}, and permits the construction of normalizable wave-packets (at least in MSS).


As in~\cite{Magueijo:2021pvq,Alexandre:2022npo,Gielen:2022dhg}, we work in connection space, so that the wave function is diagonal in the representation of a variable which, classically and in the absence of torsion, becomes $b=\dot a$. This is the inverse comoving Hubble length. It decreases (increases) for decelerating (accelerating) expansion. We will examine this feature by writing $b$ as a function of physical $T$ (rather than any coordinate $t$). 

Before embarking on the post-selected model in the next section, we first describe the general classical and quantum solutions of the base model we will use.


\subsection{The classical trajectories}

We specifically follow the minisuperspace model of~\cite{Gielen:2022dhg} for a Lambda and radiation filled Universe, using only the unimodular clock. The canonical pairs and Hamiltonian are defined from:
\begin{align}
     S&=\int dt \left (\dot b a^2+\dot \phi T -NH \right),\\
     H&=a\left[-(b^2+k)+\frac{m}{a^2}+\frac{a^2}{\phi}\right], 
\end{align}
where $\phi=3/\Lambda$ turns out to be a convenient parameterization of the $\Lambda$ term, and $m$ is the constant of motion associated with the radiation term. Variation with respect to the lapse function $N$ results in the Hamiltonian constraint $H=0$ which can be solved as: 
\begin{equation}\label{apmsolution}
a_\pm^2(b;\phi,m)=\frac{\phi}{2}\left(V(b)\pm \sqrt{V^2(b)-\frac{4m}{\phi}}\right),
\end{equation}
where $V(b)=b^2+k$. The plus branch is the accelerating branch, the minus the decelerating one. One of the Hamilton's equations gives $b=\dot a/N$, as announced. The other two Hamilton equations are:
\begin{align}
    \dot b& =
Na\left(\frac{1}{\phi}-\frac{m}{a^4}\right),\\
\dot T&=\frac{N a^3}{\phi^2},
\end{align}
i.e. the Raychaudhuri equation and the ``time formula'' for physical time $T$ conjugate to $\phi$.
We can obtain a description in terms of physical time $T$ by dividing these equations, resulting in: 
\begin{equation}
    \frac{db}{dT}=\frac{\phi^2}{a^2}\left(\frac{1}{\phi}-\frac{m}{a^4}\right)
\end{equation}
and using the solution \eqref{apmsolution} to the Hamiltonian constraint to eliminate $a$. Integrating gives us $b=b(T)$. 

In the radiation-dominated limit this simplifies to:
\begin{equation}
      \frac{db}{dT}= -\frac{\phi^2}{m^2}V^3(b)
\end{equation}
which for negligible curvature $k$ integrates to: 
\begin{equation}\label{brad}
    b_r=\left(\frac{m^2}{5\phi^2 T}\right)^{1/5}\equiv \frac{A}{T^{1/5}}
\end{equation}
(up to an arbitrary shift in $T$). Our in-state will be a semiclassical state reproducing this solution. 

The Lambda dominated solution instead gives us: 
\begin{equation}
       \frac{db}{dT}=\frac{1}{V(b)}\implies X_{CS}\equiv\frac{b^3}{3}+kb=T
\end{equation}
(again up to an arbitrary shift in $T$), where $X_{CS}$ is the Chern-Simons density reduced to minisuperspace. If $k$ is negligible we have:
\begin{equation}\label{bCS}
    b_{\rm CS}=\left(3T\right)^{1/3}. 
\end{equation}

\subsection{General solutions of the quantum theory}
This theory can be canonically quantized in the usual way. If we use the $b,T$ representation\footnote{As announced we diagonalise the connection, not the metric or the densitized inverse metric.} it amounts to replacements:
\begin{align}
    a^2&\rightarrow -i\frac{\partial}{\partial b}\\
    \phi&\rightarrow i\frac{\partial}{\partial T}
\end{align}
in the Hamiltonian constraint, which thus becomes  a non-linear Schr\"odinger-like equation in time $T$. Its general solution is:
\begin{equation}
\Psi(b,T)=\int d\phi\,A(\phi)\,
\exp\left[i\Big(P(b;\phi,m)-\phi T\Big)\right],
\end{equation}
where $P$ is the Hamilton-Jacobi function, obtained from: 
\begin{equation}
\frac{\partial P}{\partial b}=\frac{\phi}{2}\left(V(b)\pm \sqrt{V^2(b)-\frac{4m}{\phi}}\right)
\end{equation}
This integrates to $P=\phi X$, where 
\begin{equation}
    X=\frac{1}{2}\int db\, \left(V(b)\pm \sqrt{V^2(b)-\frac{4m}{\phi}}\right)
\end{equation}
is a generalization of the Chern-Simons density. 
We stress that at this stage this is not a WKB solution; rather it is an exact solution of the theory\footnote{The issue of quantizing multiple branches is very important near transitions if semiclassical solutions go there~\cite{Gielen:2022dhg}. That will not be the case here, so we just need to make sure we are on the correct branch when constructing the solution.}. 

The MSS-reduction of the Kodama state can be recognized in the above\footnote{Note that the Kodama state is both exact and a WKB solution, since the WKB expansion becomes trivial after the leading order.}.  
For $m=0$ (or if radiation is negligible), $X$ does not depend on $\phi$, and becomes the Chern-Simons density:
\begin{equation}
    X_{CS}\equiv\frac{b^3}{3}+kb
\end{equation}
with the result that the waves are non-dispersive in $X$ {\it for a general $A(\phi)$} (not necessarily semiclassical). Hence the general solution is:
\begin{equation}\label{FXT}
    \Psi(b,T)=F(X_{CS}-T)
\end{equation}
for an arbitrary $F$. This form will be useful in setting up the post-selected final state. 

\subsubsection{Semiclassical solutions}
To find semiclassical solutions, we choose a Gaussian $A(\phi)$\footnote{Note this is normalized such that $\int d\phi\, |A(\phi)|^2 = 1.$ For more details about the inner product definition see~\cite{Magueijo:2021pvq,Alexandre:2022npo,Gielen:2022dhg}.}:
\begin{equation}
A(\phi)
=
\frac{1}{(2\pi\sigma_\phi^2)^{1/4}}
\exp\!\left[-\frac{(\phi-\phi_0)^2}{4\sigma_\phi^2}\right],
\end{equation}
Expanding $P$ around the central \(\phi_0\),
\begin{equation}
P(b,\phi)
\approx
P_0
+
(\phi-\phi_0)P_1
+
\frac{1}{2}(\phi-\phi_0)^2P_2,
\end{equation}
where
\begin{equation}
P_0=P(b,\phi_0),
\qquad
P_1=\left.\partial_\phi P\right|_{\phi_0},
\qquad
P_2=\left.\partial_\phi^2P\right|_{\phi_0},
\end{equation}
and performing the Gaussian integral, 
the packet becomes
to quadratic order,
\begin{equation}
\Psi(b,T)
=
{\cal N}\,
\exp\!\left[i\phi_0\big(X(b;\phi_0)-T\big)\right]
\exp\left[
-\frac{\sigma_\phi^2(X_{\rm eff}-T)^2}
{1-2i\sigma_\phi^2 P_2}
\right],
\end{equation}
with
\begin{equation}
{\cal N}
=
\frac{(8\pi\sigma_\phi^2)^{1/4}}
{\sqrt{1-2i\sigma_\phi^2 P_2}},
\end{equation}
and: 
\begin{equation}
X_{\rm eff}
=P_1=
\left.\frac{\partial P}{\partial \phi}\right|_{\phi_0}
=
X(b;\phi_0)
+
\phi_0
\left.\frac{\partial X}{\partial\phi}\right|_{\phi_0}.
\end{equation}

As announced above, in the pure \(\Lambda\) case, \(X=X_{CS}\) does not depend on \(\phi\), so
$P(b,\phi)=\phi X_{CS}(b)$  so that $X_{\rm eff}=P_1=X_{CS}$ and  $P_2=0$, resulting in: 
\begin{equation}\label{wavepacketCS}
\Psi_{CS}(b,T)
=
(8\pi\sigma_\phi^2)^{1/4}
\exp\!\left[i\phi_0\big(X_{CS}(b)-T\big)\right]
\exp\!\left[-\sigma_\phi^2\big(X_{CS}(b)-T\big)^2\right].
\end{equation}
This is indeed of the form \eqref{FXT} and is a coherent Chern-Simons wave packet. 

In the radiation-dominated regime (\(a^4\ll m\phi\), $k$ negligible), the Hamilton--Jacobi function simplifies to
\begin{equation}
P(b,\phi)\simeq -\frac{m}{b}-\frac{m^2}{5\phi b^5},
\end{equation}
so that we can compute the relevant functions:
\begin{align}
    X_r(b;\phi)&
\simeq
-\frac{m}{\phi b}
-\frac{m^2}{5\phi^2 b^5},\\
X_{r,{\rm eff}}(b)
&=
\left.\frac{\partial}{\partial\phi}\big(\phi X_r\big)\right|_{\phi_0}
\simeq
\frac{m^2}{5\phi_0^2 b^5},\label{Xeffrad}\\
P_2&
=
\left.\partial_\phi^2(\phi X_r)\right|_{\phi_0}\simeq
-\frac{2m^2}{5\phi_0^3 b^5}.
\end{align}
The wave packet therefore is:
\begin{equation}\label{wavepacket}
\Psi_r(b,T)
=
{\cal N}_r
\exp\!\left[
i\phi_0\big(X_r(b;\phi_0)-T\big)
\right]
\exp\left[
-\frac{\sigma_\phi^2\big(X_{r,{\rm eff}}(b)-T\big)^2}
{1-2i\sigma_\phi^2P_2}
\right],
\end{equation}
with
\begin{equation}
{\cal N}_r
=
\frac{(8\pi\sigma_\phi^2)^{1/4}}
{\sqrt{1-2i\sigma_\phi^2P_2}}.
\end{equation}

\subsubsection{Properties of the semiclassical solutions}
It is easy to see that the packet's peak follows the classical trajectory in both limiting cases (as indeed in general, had we found the general solution). For the radiation-dominated packet the peak lies along $X_{r,{\rm eff}}(b)=T$, equivalent to \eqref{brad} (cf. \eqref{Xeffrad}). Likewise pure Lambda $X_{CS}=T$ is equivalent to \eqref{bCS}. 
However their width in $b$ has a rather different evolution. 

For the Chern--Simons packet this is immediate. Since the packet is Gaussian in
\(X_{\rm CS}-T\), its width in \(X_{\rm CS}\) is fixed and saturates a Heisenberg uncertainty relation between Lambda and its conjugate time:
\begin{equation}
\sigma_X^{\rm CS}=\frac{1}{2\sigma_\phi},
\end{equation}
just as any other coherent state. 
Mapping this to \(b\) gives
\begin{equation}
\sigma_b^{\rm CS}
=
\frac{\sigma_X^{\rm CS}}{|dX_{\rm CS}/db|}
=
\frac{1}{2\sigma_\phi |b^2+k|}.
\end{equation}
For negligible curvature and along the peak \(b=(3T)^{1/3}\), this scales as
\begin{equation}
\sigma_b^{\rm CS}\sim T^{-2/3}.
\end{equation}

For the radiation packet the width is read instead from \(X_{r,{\rm eff}}(b)-T\). At leading order, ignoring for the moment the dispersive correction from \(P_2\), the width in \(b\) is
\begin{equation}
\sigma_b^{r}
=
\frac{1}{2\sigma_\phi}
\left|
\frac{dX_{r,{\rm eff}}}{db}
\right|^{-1}.
\end{equation}
Using
\begin{equation}
X_{r,{\rm eff}}(b)
\simeq
\frac{m^2}{5\phi_0^2 b^5},
\end{equation}
we find
\begin{equation}
\frac{dX_{r,{\rm eff}}}{db}
\simeq
-\frac{m^2}{\phi_0^2 b^6},
\end{equation}
and hence
\begin{equation}
\sigma_b^{r}
\simeq
\frac{\phi_0^2 b^6}{2\sigma_\phi m^2}.
\end{equation}
Along the radiation peak,
\begin{equation}
b(T)\simeq
\left(\frac{m^2}{5\phi_0^2 T}\right)^{1/5},
\end{equation}
so the non-dispersive part of the width scales as
\begin{equation}
\sigma_b^{r}\sim T^{-6/5}.
\end{equation}
Including the quadratic term \(P_2\), the effective width in \(X_{r,{\rm eff}}\) is enlarged to
\begin{equation}
\sigma_{X,{\rm eff}}^2
=
\frac{1}{4\sigma_\phi^2}
+
\sigma_\phi^2 P_2^2 .
\end{equation}
Using
\begin{equation}
P_2
\simeq
-\frac{2m^2}{5\phi_0^3 b^5},
\end{equation}
and mapping again to \(b\), we obtain
\begin{equation}
\left(\sigma_b^{r}\right)^2
=
\left|
\frac{dX_{r,{\rm eff}}}{db}
\right|^{-2}
\left(
\frac{1}{4\sigma_\phi^2}
+
\sigma_\phi^2 P_2^2
\right).
\end{equation}
Since
\begin{equation}
\frac{dX_{r,{\rm eff}}}{db}
\simeq
-\frac{m^2}{\phi_0^2 b^6},
\end{equation}
this gives
\begin{equation}
\left(\sigma_b^{r}\right)^2
\simeq
\frac{\phi_0^4 b^{12}}{4\sigma_\phi^2m^4}
+
\frac{4\sigma_\phi^2}{25\phi_0^2}b^2 .
\end{equation}
The first term is the width obtained by linearly mapping the Gaussian in \(X_{r,{\rm eff}}-T\) to \(b\). The second is the genuine dispersive contribution due to the \(\phi\)-dependence of \(X_r\). At sufficiently late times the latter dominates, so that
\begin{equation}
\sigma_b^r \sim b_r(T)\sim T^{-1/5}.
\end{equation}
So the fractional uncertainty is a constant in time, $\epsilon$, which is assumed to be small:
\begin{equation}\label{epsdef}
    \frac{\sigma_b^r}{b^r}=\epsilon\ll 1.
\end{equation}
We will use this particular law for definiteness, but note that by minimally changing the clock (for example using a clock dual to $\Lambda$ rather than to $\phi$) this would change, and so would the quantitative (but not qualitative) details in what follows.

\section{ Post-selected peak and emergent acceleration}\label{Sec:PostSelectMSS}

The results of the previous section are general, and will serve as reference material in what follows. We now use them to propose a two-state, pre- and post-selected, construction.

We take as initial state a semiclassical radiation-dominated Universe, leaving out the matter epoch for simplicity. We set $\Lambda$ as close to zero as wanted (or equivalently $\phi$ as large as required), so that such a universe should never accelerate at the point we are at. We do not dwell on how a more fundamental initial state, say Hartle--Hawking or Vilenkin boundary conditions, led to it. For the purposes of this paper, the point is that the forward state is already semiclassical and dominated by regular matter, here simplified to pure radiation. 

We first consider the effects of post-selection for a generic state, to then show a concrete example.

\subsection{General peak kinematics under post-selection}
The amplitude of the forwards state is sharply peaked around the classical trajectory $b_r(T)$ with width $\sigma_r(T)$ satisfying \eqref{epsdef}, so that:
\begin{equation}
|\Psi_{\rm fwd}(b,T)|
\propto
\exp\left[
-\frac{(b-b_r(T))^2}{4\sigma_r(T)^2}
\right].
\end{equation}
We first consider an arbitrary backward (post-selected) state, $\Psi_{\rm bwd}(b,T)$ with 
amplitude $A_{\rm bwd}(b,T)$. The effective two-state amplitude at relational time $T$ is then
\begin{equation}
|\Psi_{\rm eff}(b,T)|
\propto
\exp\left[
-\frac{(b-b_r)^2}{4\sigma_r^2}
\right]
A_{\rm bwd}(b,T),
\end{equation}
with a peak satisfying: 
\begin{equation}
-\frac{b-b_r}{2\sigma_r^2}
+
\partial_b \ln A_{\rm bwd}(b,T)
=0,
\end{equation}
so that: 
\begin{equation}\label{peakequation}
b_{\rm peak}-b_r
=
2\sigma_r^2
\,
\partial_b \ln A_{\rm bwd}(b,T)
\Big|_{b=b_{\rm peak}}.
\end{equation}
This relation is exact and makes no assumption on the form of the backward state.
It shows that the evolution of the peak is determined by a competition between
the intrinsic semiclassical width of the forward packet and the logarithmic slope
of the backward amplitude.

This simplifies for a Gaussian backward amplitude,
\begin{equation}
A_{\rm bwd}(b,T)
\propto
\exp\left[
-\frac{(b-b_{\rm bwd}(T))^2}{4\sigma^2_{\rm bwd}(T)}
\right].
\end{equation}
Equation \eqref{peakequation} can then be solved explicitly, with: 
\begin{equation}\label{bpeakGauss}
b_{\rm peak}(T)
=
\frac{w_r b_r+w_{\rm bwd} b_{\rm bwd}}
{w_r+w_{\rm bwd}},
\qquad
w_r=\frac{1}{\sigma_r^2},
\qquad
w_{\rm bwd}=\frac{1}{\sigma_{\rm bwd}^2},
\end{equation}
or equivalently,
\begin{equation}
b_{\rm peak}-b_r
=
\frac{\sigma_r^2}{\sigma_r^2+\sigma_{\rm bwd}^2}
\left(b_{\rm bwd}-b_r\right).
\end{equation}
Thus the peak lies between the forward radiation trajectory and the centre preferred by the backward state, weighted by their respective sharpnesses. No small-shift approximation has been made. The peak's motion may therefore resemble neither the forwards or backwards peaks, in regions where the weighting is strongly time dependent. 

We stress that the definition of inner product influences the position of the {\it probability} peak, with a shift away from the peak of $|\psi|^2$ (see~\cite{Gielen:2022dhg}). However, as long as we are in the semiclassical regime such effects should be negligible. 

In addition we note that the phase contains important information on the central $\phi$, which does not affect the peak position, but nonetheless contains important physical information: it specifies the value of $\Lambda$ in the semiclassical Hamiltonian associated with the combined state.

\subsection{Post-selected Chern--Simons soliton}

We now specialise the backward state to a purely quantum packet of the form
\begin{equation}
A_{\rm bwd}(b,T)
=
F\!\left(X_{\rm CS}(b)-T\right),
\qquad
X_{\rm CS}(b)=\frac{b^3}{3}+kb,
\end{equation}
where we take $F$ to be a Gaussian,
\begin{equation}
F(y)
=
\frac{1}{(2\pi\sigma^2)^{1/4}}
\exp\left[
-\frac{y^2}{4\sigma^2}
\right],
\end{equation}
normalized so that $
\int dy\,|F(y)|^2 = 1$.
In contrast with a common wave packet we  choose this state to lack the central plane wave factor, with its ``beats'' inside the envelope. Hence this is a purely quantum state, without a defined $\Lambda$. It is a non-dispersive Chern--Simons soliton, whose amplitude is localised
around the curve
\begin{equation}
X_{\rm CS}(b)=T,
\end{equation}
with fixed width $\sigma_X$ in the variable $X_{\rm CS}-T$.

In principle we should insert this state into the general relation \eqref{peakequation}, resulting in a quintic equation for $b$. However, as long as the backwards peak is also sufficiently narrow, near its peak we may write \(X_{\rm CS}(b)-T \simeq
(dX_{\rm CS}/db)_{b_{\rm CS}}(b-b_{\rm CS})\), where \(X_{\rm CS}(b_{\rm CS})=T\). Thus the Chern--Simons soliton is locally a Gaussian in \(b\), with centre \(b_{\rm CS}\) and width
\begin{equation}
\sigma_{{\rm CS},b}
=
\frac{\sigma}{|dX_{\rm CS}/db|_{b_{\rm CS}}}.
\end{equation}
We may therefore use the Gaussian result \eqref{bpeakGauss}, with \(b_{\rm bwd}=b_{\rm CS}\) and \(\sigma_{\rm bwd}=\sigma_{{\rm CS},b}\). With \(k=0\), so that \(b_{\rm CS}(T)=(3T)^{1/3}\) and \(\sigma_{{\rm CS},b}=\sigma_X/b_{\rm CS}^2\), and setting  \(\sigma_r=\epsilon b_r\), we find
\begin{equation}\label{bandeta}
b_{\rm peak}
=
\frac{b_r+\eta b_{\rm CS}}{1+\eta},
\qquad
\eta
=
\frac{\sigma_r^2}{\sigma_{{\rm CS},b}^2}
=
\frac{\epsilon^2 b_r^2 b_{\rm CS}^4}{\sigma^2_X}.
\end{equation}
We are interested in a regime where the forward packet is much sharper than the backward one (\(\eta\ll1\)), but $b_{\rm CS}\gg b_r$, so that the correction to $b_r$ can be significant. Then: 
\begin{equation}
b_{\rm peak}(T)
\simeq
b_r(T)
+
\frac{\epsilon^2}{\sigma_X^2}
b^2_r(T) b^5_{\rm CS}(T) .
\end{equation}
Using \eqref{brad} and \eqref{bCS} this finally becomes:
\begin{equation}
b_{\rm peak}(T)
\simeq
A T^{-1/5}
+
\frac{\epsilon^2}{\sigma_X^2}
A^2 3^{5/3} T^{19/15}.
\end{equation}
Thus the new term grows rapidly with \(T\), whereas the radiation term decreases.
The peak location starts increasing with $T$ at
\begin{equation}\label{Tturn}
T_{\rm turn}
\simeq
\left[
\frac{\sigma_X^2}
{19\,3^{2/3}\epsilon^2 A}
\right]^{15/22},
\end{equation}
marking the onset of acceleration. 

In Appendix~\ref{consistency} we examine the consistency of the approximations made ($\eta\ll 1$ and $b_{\rm CS}\gg b_r$) at $T_{\rm turn}$, and that satisfying them does not amount to sneaking in a $\Lambda$ through the back door. In particular, we show that for a sufficiently broad backward packet and a sufficiently sharp forward one, there exists a regime in which the peak turns around ($db_{\rm peak}/dT>0$) while the forward Hamiltonian remains arbitrarily close to that of a pure radiation solution. This supports the interpretation of the effect as arising from post-selection, rather than from a classical $\Lambda$ term in the dynamics.

A classical reinterpretation of the selected peak should therefore be handled
with care. In the transition regime, the post-selected trajectory can be
locally fitted by an effective fluid with \(w\approx -1\), consistent with
current observations. This is simply a reflection of the fact that we are
probing the crossover region, where the decaying radiation contribution and
the growing post-selected correction are comparable, and an effective
description in terms of a slowly varying dark component is possible.

However, extending this interpretation beyond the transition reveals its
artificial nature. The effective equation of state must evolve rapidly, moving
away from the quasi-\(\Lambda\) behaviour and asymptoting to strongly phantom
values (\(w<-1\)). Specifically, since for a flat FRW universe filled with a perfect fluid \(p=w\rho\), one finds
asymptotically \(b(T)\propto T^{-(1+3w)/[3(3+w)]}\). Matching this behaviour
to the growing term \(b\propto T^{19/15}\) implies
\(w=-31/17\simeq -1.82\), i.e. a strongly phantom component. This is because $\eta(T)$ in \eqref{bandeta} multiplies $b_{\rm CS}$ (corresponding to pure $w=-1$)\footnote{It is only after the semiclassical regime is abandoned (and $\eta\gg 1)$ that a full cross-over to $b=b_{\rm CS}$ is achieved. Alas, this state is a soliton, without a well-defined $\Lambda$, even though its behaviour mimics a generic de Sitter $T$ dependence.}.

Thus, while a classical model can always be engineered to reproduce the
selected trajectory, it does so at the price of introducing a highly tuned,
rapidly evolving, and ultimately pathological effective fluid.

This is the direct cosmological analogue of the free-particle example discussed
earlier. In that case, a simple quantum boundary condition produces a smooth
shift of the semiclassical peak, whereas any attempt to reproduce the same
motion with a classical force leads to a complicated, time-dependent, and
physically unmotivated interaction. Likewise here, the post-selected
cosmological trajectory arises naturally from the interplay between forward
semiclassical evolution and a quantum final condition, while its classical
reinterpretation obscures this simplicity behind an unnecessarily elaborate
and implausible dynamical model.

\section{The logic behind the final state and other options}
\label{Sec:FinalState}

The point has been made with one particular final state, but it could of course
have been made with others. We now explain why the Chern--Simons soliton is a
clean choice, especially with regard to fine-tuning questions and to the
cosmological constant. We also examine alternatives. 

\subsection{The rationale behind the Chern-Simons soliton final state}

The soliton \eqref{FXT} for a Gaussian $F$ should not be confused with a semiclassical de Sitter state \eqref{wavepacketCS}. It is
localized in
$ Y=X_{\rm CS}-T $
but it has no carrier phase. Thus it is not a state with a small, large, or Planckian
value of $\Lambda$. For example, a soliton with a Gaussian $F$ centred at $Y=Y_0$ has
Fourier amplitude
\begin{equation}
        A(\phi)=
        \left({2\sigma_T^2\over \pi}\right)^{1/4}
        \exp\left[-\sigma_T^2\phi^2-i\phi Y_0\right].
\end{equation}
(The phase in $A(\phi)$ is only the translation phase fixing the centre of the
soliton in $Y$; it is not an internal beat centred on a nonzero value of
$\phi$.) Hence
\begin{equation}
        \langle \phi\rangle=0,
        \qquad
        (\Delta\phi)^2={1\over 4\sigma_T^2}.
\end{equation}
A Planckian choice $\sigma_T\sim 1$ is therefore natural if no small scale is
to be introduced in the final state\footnote{If some other microscopic scale is physically preferred, it can be used instead without changing the qualitative mechanism.}. It gives a Planckian uncertainty in the
variable conjugate to $T$, not a Planckian value of $\Lambda$. Indeed, since
\begin{equation}
        \Lambda={3\over\phi},
\end{equation}
and the distribution in $\phi$ has support at $\phi=0$, the moments of
$\Lambda$ are singular. In particular,
\begin{equation}
        \langle\Lambda^2\rangle
        =
        9\int d\phi\,{|A(\phi)|^2\over \phi^2}
\end{equation}
diverges. The final state is therefore not fine-tuned with respect to the
cosmological constant. More strongly, it has no well-defined semiclassical
cosmological constant at all. It is a purely quantum state of maximal agnosticism with respect to $\Lambda$.

The only scale in the soliton is $\sigma_T$. There remains a possible worry:
if $\sigma_T$ is Planckian, why does the post-selection not become important
immediately after the primordial Planck epoch\footnote{Obviously in this model
there is also a {\it future} quantum Planck epoch, but for the time being we must
live with the habits of language.}? The answer is that $\sigma_T$ is not by
itself the strength of the effect. What matters is the relative sharpness of
the two packets, measured by $\eta$ in (\ref{bandeta}). The forward packet has
width $\sigma_r=\epsilon b_r$, whereas the soliton has local width
$\sigma_{{\rm CS},b}=\sigma_T/b_{\rm CS}^2$. Hence the pull of the final state
on the peak is suppressed by the small semiclassical width $\epsilon$ of the
forward packet. At early times the forward packet is much sharper than the
backward one, and the conditioned peak follows the radiation trajectory. The
bias becomes important only when the effective widths have evolved enough for
the final amplitude to compete with the forward radiation packet.

Thus a Planckian $\sigma_T$ does not make the acceleration contrived. It merely
avoids introducing a small final-state scale. The observed epoch of acceleration
then fixes when this natural final-state bias becomes competitive with the
forward semiclassical packet. This depends on the sharpness $\epsilon$ of the
forward state and on the radiation scale $A$, not on a tuned value of
$\Lambda$ in the final state. In fact, as shown in Appendix~\ref{consistency},
taking the forward state closer to the $\Lambda=0$ radiation limit improves the
consistency of the approximation at the turn-around.

\subsection{Overlap between initial and final state}

It is also useful to quantify how exceptional such a final state is from the
viewpoint of the incoming semiclassical packet. The physical inner product in
this representation may be written directly in terms of the coefficient
functions appearing in the general solution,
\begin{equation}
        \langle\Psi_2|\Psi_1\rangle
        =
        \int d\phi\, A_2^*(\phi)A_1(\phi),
\end{equation}
or equivalently, in sectors where a common $X$ variable is available, as the
corresponding transform in $X$. In the present comparison the $\phi$ form is
the cleaner one, since the incoming radiation packet and the Chern--Simons
soliton are naturally expressed in different $X$ variables. Taking the incoming
packet to be centred at $\phi_0$ with width $\sigma_\phi$, and the soliton to
have width
\begin{equation}
        \sigma_{\rm sol}={1\over 2\sigma_T}
\end{equation}
in $\phi$ space, the overlap is just the overlap of two normalized Gaussians,
up to the relative translation in $Y$. Choosing this relative centre so as not
to introduce an additional suppression gives
\begin{equation}
        |\langle{\rm sol}|{\rm in}\rangle|^2
        =
        {2\sigma_\phi\sigma_{\rm sol}
        \over \sigma_\phi^2+\sigma_{\rm sol}^2}
        \exp\left[
        -{\phi_0^2\over 2(\sigma_\phi^2+\sigma_{\rm sol}^2)}
        \right].
\end{equation}
Thus, for a sharply semiclassical incoming state with $\phi_0\gg 1$ and a
Planckian soliton width, the post-selection has exponentially small overlap.
This should not be confused with a hidden tuning of a small cosmological
constant. It is the usual price of imposing an unusual final boundary condition
when judged from an initial-state-only point of view.

This point is important. If the final state were regarded as one outcome in a
lottery generated by the initial state, then smaller departures from the
unconditioned trajectory would generally have larger overlap, and the question
of why this much acceleration, rather than less, would become sharp. But that
is not the two-boundary proposal. We are not sampling final states from the
unitary evolution of the initial state, or playing a lottery among universes.
We are postulating a final boundary condition, in the same sense that standard
cosmology postulates an initial one. In that framework the overlap is not a
probability to be maximised; it is a measure of how exceptional the selected
history appears if one insists on describing it from the initial state alone.

The free-particle example illustrates the same point in miniature. If the final
slit lies on the unconditioned classical trajectory, the overlap is maximal, but
there is no anomalous acceleration. To obtain acceleration, the final condition
must be displaced, and the overlap drops. In the laboratory this merely looks
like an awkward choice of selector. In cosmology the final state is instead a
boundary condition. A small overlap is then not a defect of the mechanism, but
the expected consequence of imposing two independent boundary conditions. In
particular, simplicity at the beginning and simplicity at the end need not mean
the same thing, so there is no reason for the corresponding states to have
large overlap.

\subsection{Other options for a final state?}
Finally, let us comment on the absence of a carrier phase in our chosen final
state. If the final state carried such a phase, for example centred on some
$\phi_\infty$, the peak analysis based on the amplitudes would be essentially
unchanged. The position of the peak is governed by the product of the two
envelopes. However, the semiclassical interpretation of the resulting state
would no longer be as clean, because the phases would also add. Schematically,
one would have
\begin{equation}
        S_{\rm cond}=S_{\rm fwd}+\phi_\infty X_{\rm CS}-\phi_\infty T .
\end{equation}
The time-dependent part would simply shift the frequency conjugate to $T$. If
the forward state is centred on $\phi_0$, this gives
\begin{equation}
        \phi_{\rm cond}=\phi_0+\phi_\infty .
\end{equation}
Thus, in the limit $\phi_0\rightarrow\infty$, even a finite or Planckian
$\phi_\infty$ would not by itself generate a finite cosmological constant in
the conditioned state.

The difficulty is elsewhere. The spatial part of the phase also contributes to
the Hamilton--Jacobi data of the combined state. For a Chern--Simons phase this
contribution is proportional to $\phi_\infty X_{\rm CS}$, and therefore changes
the momentum conjugate to $b$. The resulting semiclassical state would not
simply be the radiation state selected by a final amplitude; it would carry
additional phase information inherited from the final state. Interpreting the
corresponding peak as a solution of an effective classical theory would then
mix two effects: the displacement of the peak by post-selection, and a genuine
change in the Hamilton--Jacobi data of the semiclassical state.

This is why the soliton used above is the safest choice for the present
purpose. It supplies the amplitude bias needed to move the peak, but it does
not add a carrier phase which could be mistaken for an extra term in the
semiclassical Hamiltonian. The acceleration is then forced to come from the
two-boundary conditioning itself, rather than from a de Sitter-like phase
inserted through the final state. In this restricted sense the soliton gives
the cleanest demonstration of the mechanism.

\section{Physical interpretation and discussion}\label{Sec: discussion}

There is a long history of appealing to teleology (final causes) in cosmology among philosophers, from Aristotle to Teilhard de Chardin. In scientific cosmology, the imposition of final boundary conditions also has a long history, but is mostly restricted to time-symmetric cosmological models---closed Friedmann models involving re-contraction in which the arrow of time reverses in the contracting phase \cite{Gold:1962,Hawking:1985,Hawking:1993}. Such time-symmetric cosmologies have been proposed by Gold, Wheeler, Hawking and Hartle and Gell-Mann, the latter two in the context of quantum cosmology \cite{Hartle:1983,GellMann:1994}.


The construction presented in this paper is somewhat different, because although we posit both initial and final conditions, there is no restriction to time symmetry. Rather, we appeal to a simple quantum-mechanical fact. Given an initial state, ordinary unitary evolution assigns amplitudes to a range of possible final outcomes. If, at a later time, one conditions on a particular final outcome, the statistics of intermediate quantities are no longer those obtained from the initial state alone. In the language of pre- and post-selection \cite{Aharonov1964,Aharonov2008}, the relevant intermediate object is not simply the forward-evolved state, but the product of the forward-evolved initial state and the backward-evolved final state.


In ordinary laboratory quantum mechanics this statement has a precise operational meaning. One may prepare a large ensemble of identical systems, allow them to evolve under the same Hamiltonian, and finally retain only those members for which a specified projective measurement outcome is obtained at the final time. On this post-selected subensemble, weak measurements made at intermediate times, averaged over the subensemble, do  not generally coincide with eigenvalues of the measured operator, and may in fact be quite bizarre, even apparently ``miraculous''\footnote{The term in this context has acquired a technical meaning, since the works of Aharonov and collaborators.}: spins outside their usual range, negative kinetic energies, or other values which would look impossible if interpreted as ordinary measurement outcomes \cite{Aharonov:1988,Popescu1993,Jordan:2024}. The point is not that a single weak measurement reveals such a value directly. A single weak measurement is subject to noise and is only minimally informative. Rather, the weak value is a statistical property of a selected ensemble, obtained in the large ensemble limit in which the back-reaction of the measurement on a specific, individual measured system is negligible while the ensemble average becomes sharp.


The regime of weak measurement with post-selection is a straightforward consequence of standard laboratory quantum mechanics, and has received a great deal of experimental confirmation. The central point of this paper is to extend the concept of quantum post-selection to cosmology, where the system is not an object inside the universe but, in a minisuperspace approximation, the universe itself, 
with observers reduced to irrelevant internal “microbial” activity. We postulate an initial semiclassical cosmological wavefunction and a final quantum state. Between them, the effective semiclassical history is determined by the two-boundary amplitude. As in the single-particle illustration (section~\ref{Freeparticel}), the resulting peak in minisuperspace need not follow the classical trajectory associated with the forward state alone (section~\ref{Sec:PostSelectMSS}). In the model studied above, the forward state Hamiltonian corresponds to a radiation-dominated semiclassical universe with vanishing cosmological constant, while the final state is chosen to be a normalizable Chern–Simons soliton. Their product has a peak which first follows the trajectory of a radiation-dominated Friedmann universe, but later departs from it and enters an accelerating regime, mimicking the effect of dark energy coming to dominate.

This acceleration is therefore not produced by adding a cosmological constant, a dark-energy fluid, a modified gravitational equation, or a new local force. It is produced simply by conditioning on a final quantum state. The analogy with the free-particle example in Section II is direct. A free particle governed by a free Hamiltonian can acquire an accelerated conditional peak once a final state is imposed (in our case by adding a single slit screen and retaining only those runs that clicked). A classical observer attempting to explain that peak would infer a complicated and unmotivated force. Likewise, a classical cosmologist attempting to reproduce the post-selected cosmological trajectory can always introduce an effective fluid or an effective Friedmann equation, but the resulting component is contrived. Near the transition it may resemble $w = -1$, as observations suggest, but its extrapolation is not that of a natural local source. In this sense the classical reconstruction is a useful diagnostic, not necessarily the explanation.

There is, however, an important interpretational distinction between the laboratory and cosmological cases. In the laboratory one may speak literally of an ensemble of identically prepared systems and of a final projective measurement used to define the post-selected subset. Cosmology does not obviously provide such an ensemble, nor an external observer performing a final measurement on the wavefunction of the universe. Thus the post-selected final state cannot be understood as an experimental filtering procedure applied from outside the universe. It must instead be regarded as a final boundary condition, on the same logical footing as the initial condition.

What justification can be given for cosmological post-selection? First note that standard cosmology assumes a dynamical law (e.g. Einstein’s general theory of relativity), but doesn’t demand any specific initial condition. There is no “law of initial conditions”, only an appeal to philosophical criteria, such as “naturalness”, for the choice. A widely-accepted “naturalness” approach is to assume that the universe started out simple, and evolved greater complexity over time. The Hartle-Hawking no-boundary proposal~\cite{Hartle:1983} is one example of this philosophy. Although there has been disagreement over the past century of precisely what form the cosmological initial conditions should take~\cite{Hartle:1983,Vilenkin:1984,Linde:1984}, there has been little dissent about the need to assign some form of primordial state. Part of the reason it seems natural to pick an initial state of the universe is that in human experience we feel free to set up a physical system at some moment and then follow its evolution over time. We are not free to establish a final condition unless it occurs as the unique evolution from an initial state. So there is a psychological prejudice in favour of fixing initial but not final states. But that is in the realm of human affairs and limitations. When it comes to the universe, there is no justification for being biased toward initial rather than final conditions. And if we suppose that the initial state of the universe is not arbitrary but special in some sense, then we can with equal justification say the same about the final state. If the criteria were the same for both, e.g. simplicity, then it will generally not be the case that the final quantum state is merely the unitary evolution of the initial state, but would be a branch thereof, and possibly one with a very small amplitude, in which case the wave function at intermediate times is able to display peculiar “miraculous” behaviour.

A dramatic and simple illustration is provided by a black hole in a Friedmann--Robertson--Walker universe~\cite{Hawking:1975vcx,Unruh:1976db,Birrell:1982ix}. In the standard treatment, the initial state of the field is chosen to be the Unruh vacuum, suitably generalized to the cosmological setting, on grounds of simplicity. The corresponding final state contains the outgoing Hawking flux. Equivalently, the in-vacuum, when examined in the out-region by an inertial detector, contains particles with an approximately thermal spectrum. In an ever-expanding universe this flux is redshifted to arbitrarily low temperature, but it never becomes a strict vacuum. Suppose, however, that one imposes a strict out-vacuum as the final state. In the far future this seems like a negligible adjustment: one has replaced an arbitrarily cold state by the vacuum. Propagated backwards, however, the change is enormous. Near the past black-hole horizon the state is no longer regular in the way required by the Unruh construction; the stress tensor diverges on the horizon, and its back-reaction would invalidate the classical black-hole geometry. Thus a physically tiny-looking change in the far future can have dramatic consequences in the past. This is the sense in which the effect is a quantum miracle. It follows merely from imposing a simple quantum state at both ends. There is no fundamental reason to regard simplicity of the initial state as legitimate while refusing the same criterion for the final state. Yet imposing simplicity at both ends can have momentous implications in the intermediate region.

From this viewpoint, the apparent improbability of the selected final branch can be interpreted in two ways. If one insists on an initial-state-only formulation, then a nearly orthogonal final state is indeed a low-probability outcome. The explanation of acceleration would then be the result of a huge fluke,
 and one might need to appeal to an anthropic or multiverse-type argument to make sense of why our universe lies on such an improbable branch~\cite{Weinberg:1987,Barrow:2002,BarrowTipler:1986}.
 Alternatively, one may take the two-boundary formulation seriously and postulate both an initial and a final simplicity condition as part of a more general cosmological law. Then the question would not be why an unlikely final state was obtained by chance, but why the universe is indeed subject to simple boundary conditions at both ends.

This second attitude is less radical than it may first appear. As explained above, we are accustomed to specifying simple initial conditions for the universe, even though such conditions are not derived from the local equations of motion. A final boundary condition is conceptually similar, though less familiar. In the present model it has a particularly economical consequence: it produces an intermediate semiclassical history which appears to accelerate, without inserting a small fictitious force, such as derives from a cosmological constant, into the semiclassical description at the intermediate stage. The acceleration is not the result of a finely tuned local vacuum energy, but of a global quantum boundary condition.

The proposal also clarifies the role of observation. Cosmological measurements such as super-novae, BAO, or the CMB are not active projective measurements repeatedly performed on an ensemble of universes. They are passive records within an already semiclassical history: they are “weak” in the sense of the quantum theory of weak values. They do not generate the post-selection; they reveal the post-selected branch. Thus the mechanism we propose in this paper is not spoiled by observation, provided the branch has decohered sufficiently to admit a semiclassical description~\cite{GellMann:1994,Zurek:2003}. In this respect the weak-measurement analogy is useful but should not be over-literalized. Its purpose is to show that anomalous intermediate behaviour conditioned on a final state is a genuine quantum effect, not that cosmological observations are laboratory weak measurements.

Several technical questions remain open. The first concerns the physical status of the final state. We have used a normalizable Chern–Simons soliton because it supplies an amplitude bias without imposing internal semiclassical beats corresponding to a nonzero cosmological constant. The forward state therefore retains the phase information associated with $\Lambda\rightarrow 0$, while the backward state reshapes the amplitude and hence the peak.

More general final states, including states closer to ordinary Chern--Simons/Kodama or Hartle--Hawking wave packets~~\cite{Hartle:1983,Kodama:1988,Smolin:2002,Magueijo:2020ugp,Alexander:2022ocp},
may lead to similar peak motion, 
but their phases must be treated carefully: the phases add, and this affects the semiclassical Hamilton–Jacobi data of the combined state. This will be investigated in future work.

A second question concerns fluctuations around the selected peak. A purely self-dual de Sitter final state would be too rigid, while a more general final state might carry gravitational fluctuations, perhaps including tensor modes. The model therefore suggests possible observational discriminants, not only through the background expansion history but also through the pattern of perturbations inherited from the final boundary condition. This remains to be developed.

The broader lesson is that an apparently dynamical cosmic acceleration need not originate in a local dynamical source. It may instead be the semiclassical shadow of a quantum boundary condition. In an initial-value formulation this looks miraculous, just as anomalous weak values look miraculous when described without reference to post-selection. In a two-boundary formulation, however, the miracle is only apparent: the intermediate history is constrained by both ends.

\section*{Acknowledgments} We thank the Caribbean Future of Science Symposium, where this work was started out of a Q\&A misunderstanding. We further thank Pauline Davies, Flaminia Giacomini, Andrew Jordan, Laura Mersini, Ian Moss and Sandu Popescu for comments and discussions. JM is partly supported by STFC Consolidated Grant ST/T000791/1.


\appendix

\section{Acceleration and spreading}\label{Ap:details1}

We give here the explicit acceleration formula following from the peak trajectory in Section II. Write
\begin{equation}
x_{\rm cl}(t)=x_0+vt,
\qquad
v=\frac{p_0}{m},
\end{equation}
and set
\begin{equation}
\tau=t_f-t,
\qquad
x_{\rm cl}(t)=x_{\rm cl}(t_f)-v\tau .
\end{equation}
It is useful to define
\begin{equation}
\Delta:=X_f-x_{\rm cl}(t_f),
\qquad
\beta_0:=\frac{\hbar^2}{4m^2\Sigma^2},
\qquad
D(\tau):=\sigma^2+\Sigma^2+\beta_0\tau^2 .
\end{equation}
Then, using
\begin{equation}
W^2(\tau)=\Sigma^2+\beta_0\tau^2,
\end{equation}
the shift of the peak may be written as
\begin{equation}
\delta X(\tau)
=
\frac{\sigma^2\big(\Delta+v\tau\big)}{D(\tau)} .
\end{equation}
Since \(d/dt=-d/d\tau\), the peak velocity and acceleration are
\begin{equation}
\dot X(t)=v-\frac{d}{d\tau}\delta X(\tau),
\qquad
\ddot X(t)=\frac{d^2}{d\tau^2}\delta X(\tau).
\end{equation}
A straightforward calculation gives
\begin{equation}\label{accelgeneral}
\ddot X(t)
=
\frac{2\sigma^2\beta_0}{D(\tau)^3}
\left[
\beta_0 v\tau^3
+
3\beta_0\Delta\tau^2
-
3v(\sigma^2+\Sigma^2)\tau
-
\Delta(\sigma^2+\Sigma^2)
\right].
\end{equation}
In particular, near the final time,
\begin{equation}
\ddot X(t_f)
=
-\frac{2\sigma^2\beta_0\Delta}{(\sigma^2+\Sigma^2)^2}.
\end{equation}
Thus the sign of the acceleration is controlled by the mismatch between the final boundary condition and the classical endpoint: if \(\Delta<0\) the peak accelerates, while if \(\Delta>0\) it decelerates. Equivalently, the peak is pulled towards the location preferred by the final state. Schematically,
\begin{equation}
\ddot X\sim -\kappa (X_f-X),
\qquad
\kappa\sim
\frac{\sigma^2\beta_0}{(\sigma^2+\Sigma^2)^2},
\end{equation}
so the effect scales as
\begin{equation}
\ddot X\propto
\frac{\hbar^2}{m^2}
\frac{\sigma^2}{\Sigma^2(\sigma^2+\Sigma^2)^2},
\end{equation}
and disappears in the classical limit \(\hbar\to0\).

We have so far kept the forward width fixed. Including spreading of the forward packet amounts to replacing
\begin{equation}
\sigma^2
\longrightarrow
\sigma^2(t)
=
\sigma^2+\frac{\hbar^2 t^2}{4m^2\sigma^2}
\end{equation}
in the above formulae, while the backward width remains
\begin{equation}
W^2(\tau)
=
\Sigma^2+\frac{\hbar^2\tau^2}{4m^2\Sigma^2}.
\end{equation}
The spreading time is inversely proportional to the initial width squared. Hence, for \(\Sigma\ll\sigma\),
\begin{equation}
t_{\rm spread}^{(f)}
\sim
\frac{2m\Sigma^2}{\hbar}
\ll
t_{\rm spread}^{(i)}
\sim
\frac{2m\sigma^2}{\hbar}.
\end{equation}
The backward packet therefore becomes broad much earlier than the forward one. This creates an extended regime in which the forward packet remains narrow and semiclassical, while the backward packet already influences the peak. The correction to the classical trajectory becomes important when \(W(\tau)\sim\sigma\), namely
\begin{equation}
\tau_{\rm corr}\sim \frac{2m\sigma\Sigma}{\hbar}.
\end{equation}

\section{Further details on the free particle}\label{Ap:details2}

This appendix collects three checks on the free-particle example which are useful but not needed in the main text. First, we show that the conditioned packet can remain semiclassical even when the displacement of its peak becomes significant. Second, we check that the phase of the two-boundary state encodes the momentum of the selected peak, so that the amplitude motion is consistent with the Hamilton--Jacobi data. Third, we spell out the energy accounting: the post-selected subensemble may look as if a force has done work, but the effect is paid for by the exponentially small probability of successful post-selection.

\subsection{Semiclassicality of the conditioned packet}

We now examine whether the conditioned packet remains semiclassical in the regime where the deviation of the peak from the classical trajectory becomes significant. From (B1), corrections become relevant when $W(\tau)\lesssim\sigma$, so that the weighting factor $\alpha(\tau)=\sigma^2/(\sigma^2+W^2)$ is no longer small. At the crossover $W\sim\sigma$, one has $\alpha\simeq 1/2$, so the peak has already moved by an $O(1)$ fraction of $(X_f-x_{\rm cl})$.

In this regime the effective width of the combined packet is
\begin{equation}
\sigma_{\rm eff}^2=\frac{\sigma^2 W^2}{\sigma^2+W^2},
\end{equation}
so that $\sigma_{\rm eff}\simeq \sigma/\sqrt{2}$ at $W\sim\sigma$. Thus the position-space localization is essentially unchanged from that of the original semiclassical packet.

The nontrivial issue is the momentum spread. Ignoring for the moment the phase-induced chirp, one has $\Delta p_{\rm eff}\simeq \hbar/(2\sigma_{\rm eff})$, so that at the crossover $\Delta p_{\rm eff}\simeq \sqrt{2}\,\hbar/(2\sigma)$. Hence, if the original packet satisfied $\sigma p_0\gg\hbar$, the conditioned packet remains semiclassical up to factors of order unity.

The backward evolution generates a quadratic phase, contributing an additional ``chirp'' to the momentum variance. Near the scale $\tau_{\rm corr}\sim 2m\sigma\Sigma/\hbar$ this contribution scales as $\Delta p_{\rm chirp}\sim \hbar\,\Sigma/\sigma^2$, and is therefore subdominant provided $\Sigma\ll\sigma$. In this regime the momentum uncertainty remains controlled by the usual Heisenberg term.

We conclude that there exists a window $\Sigma\ll W(\tau)\lesssim\sigma$, with $\sigma p_0\gg\hbar$, in which the peak deviation is significant while the state remains semiclassical in both position and momentum. Only for $W\ll\sigma$ does the momentum spread become parametrically large and the state depart from the semiclassical regime.

\subsection{Phase of the backward packet and consistency of the peak momentum}

Although the final state at $t_f$ may be chosen real (e.g.\ a Gaussian centered at $X_f$), backward Schrödinger evolution necessarily generates a non-trivial phase. For a free particle one has
\begin{equation}
\tilde\psi_f(x,t)\;=\;\int dx'\,K(x,t;x',t_f)\,\psi_f(x',t_f),
\end{equation}
with propagator
\begin{equation}
K \sim \exp\!\left[\frac{i}{\hbar}S_{\rm cl}(x,t;x',t_f)\right],\qquad
S_{\rm cl}=\frac{m(x-x')^2}{2(t-t_f)}.
\end{equation}
For $t<t_f$ this yields a Gaussian with broadened width and a quadratic phase,
\begin{equation}
\tilde\psi_f(x,t)\;\propto\;
\exp\!\left[-\frac{(x-X_f)^2}{4W^2(\tau)}\right]\,
\exp\!\left[\frac{i}{\hbar}S_f(x,t)\right],
\end{equation}
where $\tau=t_f-t$ and, up to irrelevant constants,
\begin{equation}
S_f(x,t)\;\approx\;-\frac{m(x-X_f)^2}{2\tau}\,\frac{\Sigma^2}{W^2(\tau)}.
\end{equation}
Thus the backward-evolved packet acquires a phase gradient
\begin{equation}
\partial_x S_f(x,t)\;\approx\;\frac{m(X_f-x)}{\tau}\,\frac{\Sigma^2}{W^2(\tau)},
\end{equation}
which corresponds to the momentum required to reach $X_f$ at $t_f$.

The conditioned state is the product
\begin{equation}
\Psi_{\rm cond}(x,t)\;\propto\;A_i(x,t)A_f(x,t)\,
\exp\!\left[\frac{i}{\hbar}(S_i+S_f)\right].
\end{equation}
While the amplitudes $A_iA_f$ determine the position of the peak $X(t)$, the summed phase defines a local momentum field
\begin{equation}
p(x,t)=\partial_x(S_i+S_f).
\end{equation}
In the semiclassical regime (narrow packet, slowly varying phase), the peak follows this momentum:
\begin{equation}
m\,\dot X(t)\;\approx\;\partial_x(S_i+S_f)\big|_{x=X(t)}.
\end{equation}
Hence the trajectory selected by the amplitude product is dynamically consistent with the phase of the two-boundary state: the peak moves with the momentum encoded in the summed Hamilton--Jacobi phases. The situation is quite different in quantum cosmology with a Chern-Simons soliton as a final state. 

\subsection{Energy accounting}\label{energyaccounting}

One might wonder whether the free-particle version of the effect could be used,
at least in principle, as an energy source. After all, by conditioning on a
suitable final state one may select intermediate histories whose peak appears
to accelerate, and a classical observer would be tempted to attribute this to a
force doing work. There is no paradox here. The apparent work is a property of
the post-selected subensemble, not of the full ensemble.

The price is the success probability of the post-selection. If the final
selector is centred a distance
$
        \Delta=X_f-x_{\rm cl}(t_f)$
from the unconditioned classical endpoint, then, for Gaussian packets, the
success probability is of order
\begin{equation}
        p_{\rm succ}\sim
        \exp\left[-{\Delta^2\over 2(\sigma_f^2+\Sigma^2)}\right],
\end{equation}
up to the usual Gaussian prefactor. Thus the number of preparations required
per successful run scales as
\begin{equation}
        N_{\rm runs}\sim {1\over p_{\rm succ}}
        \sim
        \exp\left[{\Delta^2\over 2(\sigma_f^2+\Sigma^2)}\right].
\end{equation}
Large anomalous accelerations therefore come at the cost of exponentially rare
post-selection. Any attempt to extract a finite amount of work from the
selected particles would have to pay this price in discarded runs, preparation,
selection and readout. The full ensemble remains governed by the free
Hamiltonian; only the conditioned subensemble looks as if an external force has
acted. 
In this sense post-selection can act as an energy source for the selected
subensemble, but not as a free source in the unconditioned theory. The effect is
real conditionally, but paid for statistically.

\section{Consistency of the approximations at the turning point}\label{consistency}

The turning time obtained in \eqref{Tturn},
\begin{equation}
T_{\rm turn}\simeq
\left[
\frac{\sigma_X^2}{19\,3^{2/3}\epsilon^2 A}
\right]^{15/22},
\qquad
A=\left(\frac{m^2}{5\phi_0^2}\right)^{1/5},
\end{equation}
must lie within the regime where the approximations used to derive it are self-consistent ($\eta\ll1$ and $b_{\rm CS}\gg b_r$). 
The ratio of the two classical solutions is
\begin{equation}
\frac{b_{\rm CS}}{b_r}
=
\frac{3^{1/3}T^{8/15}}{A},
\end{equation}
so that at the turning point
\begin{equation}
R_{\rm turn}:=
\frac{b_{\rm CS}}{b_r}\bigg|_{T_{\rm turn}}
\sim
\left(\frac{\sigma_X}{\epsilon}\right)^{8/11}
A^{-15/11}.
\end{equation}
Hence the condition $b_{\rm CS}\gg b_r$ at $T_{\rm turn}$ requires
\begin{equation}
\frac{\sigma_X}{\epsilon}\gg A^{15/8}.
\end{equation}
On the other hand, the small-$\eta$ condition at the turning point follows from the general relation
\begin{equation}
\eta_{\rm turn}=\frac{3}{19}\frac{b_r}{b_{\rm CS}},
\end{equation}
so that $b_{\rm CS}\gg b_r$ automatically implies $\eta_{\rm turn}\ll1$.

An important feature of this result is that the hierarchy at $T_{\rm turn}$ is controlled not only by the background parameters $(m,\phi_0)$, but also by the ratio $\sigma_X/\epsilon$, i.e. by the relative widths of the backward and forward packets. In particular, taking $\Lambda\to0$ (i.e. $\phi_0\to\infty$) sends $A\to0$, and therefore
\begin{equation}
R_{\rm turn}\to\infty,
\qquad
\eta_{\rm turn}\to0,
\end{equation}
so that both conditions become parametrically better satisfied.
Thus, for a sufficiently broad backward packet and a sufficiently sharp forward one, there exists a regime in which the peak turns around ($db_{\rm peak}/dT>0$) while the forward Hamiltonian remains arbitrarily close to that of a pure radiation solution.

\end{document}